\documentclass[aps,twocolumn,superscriptaddress]{revtex4}
\usepackage{epsfig}
\usepackage{times}
\begin{document}

\title{Leaders should not be conformists in evolutionary social dilemmas}

\author{Attila Szolnoki}
\email{szolnoki@mfa.kfki.hu}
\affiliation{Institute of Technical Physics and Materials Science, Research Centre for Natural Sciences, Hungarian Academy of Sciences, P.O. Box 49, H-1525 Budapest, Hungary}

\author{Matja{\v z} Perc}
\email{matjaz.perc@uni-mb.si}
\affiliation{Faculty of Natural Sciences and Mathematics, University of Maribor, Koro{\v s}ka cesta 160, SI-2000 Maribor, Slovenia}
\affiliation{CAMTP -- Center for Applied Mathematics and Theoretical Physics, University of Maribor, Krekova 2, SI-2000 Maribor, Slovenia}

\begin{abstract}
The most common assumption in evolutionary game theory is that players should adopt a strategy that warrants the highest payoff. However, recent studies indicate that the spatial selection for cooperation is enhanced if an appropriate fraction of the population chooses the most common rather than the most profitable strategy within the interaction range. Such conformity might be due to herding instincts or crowd behavior in humans and social animals. In a heterogeneous population where individuals differ in their degree, collective influence, or other traits, an unanswered question remains who should conform. Selecting conformists randomly is the simplest choice, but it is neither a realistic nor the optimal one. We show that, regardless of the source of heterogeneity and game parametrization, socially the most favorable outcomes emerge if the masses conform. On the other hand, forcing leaders to conform significantly hinders the constructive interplay between heterogeneity and coordination, leading to evolutionary outcomes that are worse still than if conformists were chosen randomly. We conclude that leaders must be able to create a following for network reciprocity to be optimally augmented by conformity. In the opposite case, when leaders are castrated and made to follow, the failure of coordination impairs the evolution of cooperation.
\end{abstract}


\maketitle

Unlike interactions among bacteria, plants or viruses \cite{turner_n99, griffin_n04, kerr_n06, biernaskie_prsb10}, interactions among humans and social animals are often driven not solely, or even not at all, by a desire to maximize fitness, but rather with an aim to socialize, to fit in, or to identify oneself with a group of like-minded individuals \cite{fiske_09, bentley_11}. This is somewhat at odds with the main determinant of success in the realm of evolutionary games \cite{hofbauer_98, mestertong_01, nowak_06, sigmund_10}, where fitness, most often quantified by a scalar payoff, reigns supreme as the key target to maximize. To take this into account, conformity has been considered as an alternative to straightforward payoff maximization, the main idea being to adopt not the strategy that is the most profitable, but the strategy that is the most common in the group or within ones interaction range \cite{wooders_geb06, pena_pre09, wang_xw_sc12, cui_pb_pa13, cui_pb_jtb14, szolnoki_rsif15}.

Since social interactions are traditionally described by models entailing networks \cite{castellano_rmp09, havlin_pst12, rand_tcs13, helbing_n13, kivela_jcn14}, evolutionary games on networks and structured populations in general have received ample attention in the recent past \cite{szabo_pr07, perc_bs10, perc_jrsi13, wang_z_epjb15}. This avenue of research owes its heyday to network reciprocity, which was discovered by Nowak and May \cite{nowak_n92b}, and has since been promoted as one of the key mechanisms for cooperation in evolutionary social dilemmas \cite{nowak_s06}. Follow-up research, mainly in the realm of statistical physics, has focused on small-world \cite{abramson_pre01}, scale-free \cite{santos_prl05}, coevolving \cite{ebel_pre02, zimmermann_pre04}, hierarchical \cite{lee_s_prl11}, bipartite \cite{gomez-gardenes_c11}, and in the past couple of years also on multilayer networks \cite{gomez-gardenes_srep12, wang_z_epl12, wang_z_srep13}. Mechanical Turk \cite{rand_jtb12} and related advances in online recruitment and large-scale data analysis have also paved the way towards social experiments to test various theoretical predictions \cite{rand_pnas11, gracia-lazaro_pnas12, rand_pnas14}.

Here we wish to extend the scope of evolutionary games \cite{rapoport_70, isaac_pc84, masuda_pla03, gunnthorsdotti_jebopa07, suri_pone11, capraro_srep14} in structured populations by considering social dilemmas, and notably the prisoner's dilemma game as the most commonly used workhorse to that effect \cite{fudenberg_e86, nowak_n93, imhof_pnas05, fu_epjb07, cheng_hy_njp11, fu_pre08b, rong_pre10, laird_ijbc12, fu_pre09, antonioni_pone11, dai_ql_njp10, tanimoto_pre12, shigaki_epl12, sun_jt_njp10, hilbe_pnas13, wu_t_pa14, rong_epl13, wu_js_pa14}, with randomly assigned as well as player-specific conformity. As noted previously, this entails the adoption of the strategy that is most common within the interaction range of the player, regardless of the expected payoff \cite{henrich_ehb98}. By adopting the most common strategy, conformists coordinate their behavior in a way that minimizes individual risk and ensures that their payoff will not be much lower than average. Importantly, a recent economic experiment involving the public goods game with institutionalized incentives has confirmed such behavior \cite{wu_jj_srep14}. In addition to randomly assigned conformity \cite{szolnoki_rsif15}, we consider degree, collective influence \cite{morone_n15}, as well as teaching activity \cite{szolnoki_epl07} as criteria to determine which players conform and which aspire to payoff maximization. As we will show, regardless of which criterion is applied, for socially desirable outcomes with ample cooperation leaders should not be conformists.

\section*{Results}

\subsection*{Evolutionary games with conformists}
We consider evolutionary social dilemmas on either the scale-free or the Erd{\H os}-R{\'e}nyi random network (see Methods for details), or on interaction networks with a uniform degree distributions, such as the random regular graph and the square lattice, where each player $x$ is initially designated either as cooperator ($C$) or defector ($D$) with equal probability.
Each instance of the game involves a pairwise interaction where mutual cooperation yields the reward $R$, mutual defection leads to punishment $P$, and the mixed choice gives the cooperator the sucker's payoff $S$ and the defector the temptation $T$. We predominantly consider the weak prisoner's dilemma, such that $T>1$, $R=1$ and $P=S=0$, but we also consider the true prisoner's dilemma where $S<0$.

The standard Monte Carlo simulation procedure comprises the following elementary steps. First, according to a random sequential update protocol, a randomly selected player $x$ acquires its payoff $\Pi_x$ by playing the game with all its neighbors. Next, player $x$ randomly chooses one neighbor $y$, who then also acquires its payoff $\Pi_y$ in the same way as previously player $x$. To avoid payoff-related effects that are due to heterogeneous interaction topologies \cite{santos_prl05}, we normalize the payoff with the degree of the corresponding player $\widetilde{\Pi}_x = \Pi_x/k_x$. After both players acquire their payoffs, player $x$ adopts the strategy $s_y$ from player $y$ with a probability determined by the Fermi function
\begin{equation}
\Gamma(\widetilde{\Pi}_x - \widetilde{\Pi}_y)=\frac{1}{1+\exp((\widetilde{\Pi}_x-\widetilde{\Pi}_y)/K)}\,\,,
\label{fermi}
\end{equation}
where $K=0.1$ quantifies the uncertainty related to the strategy adoption process \cite{blume_l_geb93, szabo_pr07}. In agreement with previous works, the selected value ensures that strategies of better-performing players are readily adopted by their neighbors, although adopting the strategy of a player that performs worse is also possible \cite{perc_pre08b, szolnoki_njp08}. This accounts for imperfect information, errors in the evaluation of the opponent, and similar unpredictable factors.

To introduce conformity, we designate a fraction $\rho$ of the population as being conformity-driven, which influences the strategy adoption rule. Instead of Eq.~\ref{fermi}, if player $x$ is a conformist, we use
\begin{equation}
\Gamma(N_{s_x} - k_h)= \frac{1}{1+\exp((N_{s_x}-k_h)/K)}\,\,,
\label{conform}
\end{equation}
where $N_{s_x}$ is the number of players adopting strategy $s_x$ within the interaction range of player $x$, while $k_h$ is one half of the degree of player $x$. By using Eq.~\ref{conform}, player $x$ is most likely to adopt the strategy that is, at the time, the most common in its neighborhood. As in Eq.~\ref{fermi}, here too $K=0.1$ introduces some uncertainty to the process, such that it is not completely impossible for a conformist to adopt a strategy that is in the minority among its neighbors. If, however, the number of cooperators and defectors in the neighborhood is equal, the conformity-driven player will change its strategy with probability $1/2$. We also emphasize that, in this study, a conformist uses only local information. In particular, a conformity-driven player simply tends to follow the majority in its local neighborhood. An alternative approach might entail a conformist having available global information about the whole population and act accordingly. In such a case, however, the global information may likely suppress the vital importance of local information stemming from the direct neighbors. Due to several open questions and options on implementation, we leave it to future studies to determine the merits of global conformity.

A key consideration in this paper is which players make up the fraction $\rho$ of the population that act as conformists. We consider and compare several different options. The simplest option is to assign conformists uniformly at random, which we use as the benchmark case. Secondly, on heterogeneous interaction networks, such as scale-free and Erd{\H os}-R{\'e}nyi random networks, we use the degree $k_x$ of each player, whereby the probability to be designated as a conformist can be either $k_x/k_{max}$ or $1-k_x/k_{max}$ for degree-related or inversely-degree-related assignment, respectively. In the former case, high-degree players, i.e., the hubs or leaders, most certainly end up as conformists, while in the latter case low-degree players, i.e., the masses or the periphery, are the most likely conformists. Thirdly, again of relevance on heterogeneous networks, we use the concept of collective influence of players to distinguish them \cite{morone_n15}. In particular, we calculate the collective influence $CI_x$ at depth $1$ for every player $x$ (see \cite{morone_n15} for details), where the maximal value is $CI_{max}$. The incentive for player $x$ to act as a conformist is then simply based on the ratio $CI_x/CI_{max}$.
Accordingly, players with the highest collective influence at depth $1$ in the network will most likely end up using Eq.~\ref{conform} rather than Eq.~\ref{fermi} for the strategy adoption. In principle, this approach is the same as the degree-related assignment of conformity, in that the leaders are the likeliest to conform. Lastly, we check our observations on homogeneous networks, where all players have an identical number of neighbors. Instead of considering topological differences, we thus assign a different strategy pass capacity $w_x$ to each player as a pre-factor in  Eq.~\ref{fermi} \cite{szolnoki_epl07}, such that the distribution is $P(w) \propto w^\frac{1}{2}$, where $w_x \in [0.01,1]$ interval. This practically means that the majority of the players has a very low $w_x$ value, while only a few have $w_x \approx 1$ (the distribution is shown in the inset of Fig.~\ref{SQR}). As with the degree, we consider $w$-related (the probability is simply equal to $w_x$ because $w_{max}=1$) or inversely-$w$-related (1-$w_x$) assignment of conformity, such that the leaders or the masses are preferentially considered as conformists, respectively.

\subsection*{Evolutionary dynamics}

\begin{figure}
\centerline{\epsfig{file=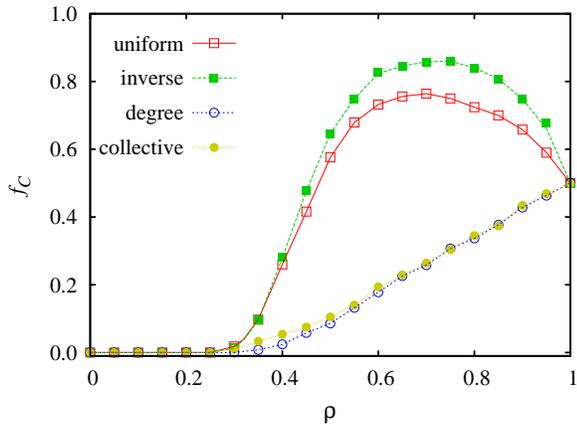,width=8.5cm}}
\caption{Fraction of cooperators $f_C$ in dependence on the fraction $\rho$ of the population that act as conformists, as obtained on the scale-free network for different selection rules indicated in the legend. Assigning the conformist status to players with the highest degree (degree) or collective influence (collective) in the scale-free network strongly impairs the constructive interplay between heterogeneity and coordination, thus leading to lower levels of cooperation than are obtained if the same status is assigned randomly (uniform) or to low-ranking players (inverse). Scale-free networks with $N=10^4$ nodes and the weak prisoner's dilemma $(T,S)=(1.1,0)$ parametrization have been used. Presented results are averages over $5000$ independent realizations.}
\label{SF}
\end{figure}

We begin by presenting the main results on scale-free networks in Fig.~\ref{SF}, where we show how the fraction of cooperators ($f_C$) in the stationary state depends on the fraction $\rho$ of the whole population that is made up of conformist players. As stated in the Introduction, these players do not aspire to maximal payoffs as is traditionally assumed in evolutionary games, but rather, they prefer to adopt the most common strategy within their interaction range. As we have reported in \cite{szolnoki_rsif15}, when conformity is assigned randomly and to a sufficiently high fraction of the population, the spatial selection for cooperation is enhanced, and the level of cooperation in the stationary state is higher than in the absence of conformist players. Results presented in Fig.~\ref{SF} show, however, that this effect can be either enhanced or destroyed based on the preference of who will be made to conform. Specifically, if conformists are players with a high degree or high collective influence in the network, the evolution of cooperation is significantly impaired. The constructive interplay between heterogeneity and coordination is almost completely lost. Naturally, we can still observe a gradual increase of $f_C$ as the fraction of conformists increases, but this is just a simple consequence of the fact that we reach a strategy neutral state at $\rho=1$.

\begin{figure}
\centerline{\epsfig{file=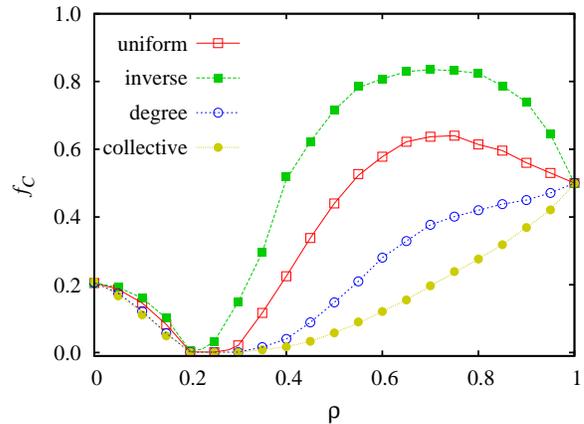,width=8.5cm}}
\caption{Fraction of cooperators $f_C$ in dependence on the fraction $\rho$ of the population that act as conformists, as obtained on the Erd{\H os}-R{\'e}nyi random network for different selection rules indicated in the legend. As for the scale-free network in Fig.~\ref{SF}, here assigning the conformist status to players with the highest degree (degree) or collective influence (collective) also leads to lower levels of cooperation than are obtained if the same status is assigned randomly (uniform) or to low-ranking players (inverse). Erd{\H os}-R{\'e}nyi random networks with $N=10^4$ nodes and the weak prisoner's dilemma $(T,S)=(1.1,0)$ parametrization have been used. Presented results are averages over $1000$ independent realizations.}
\label{ER}
\end{figure}

On the other hand, if conformists are preferentially selected from low-degree players, i.e., from
the masses, then an even higher level of cooperation in the stationary state is attainable, indicating an optimization of the aforementioned interplay. This result is intuitive and understandable, because high-degree players are naturally in a position to influence a large neighborhood, and the effectiveness of this influence is facilitated further if the neighborhood is made up predominantly of conformist players. This interplay of heterogeneity (the ability of a small number of select players, i.e., the leaders, having a wide influence) and coordination (the pressure of conformity to select the most common strategy, even if it is not payoff-maximizing) ultimately leads to large homogenous groups or clusters competing in the population, which in the long-run reveals the benefits of cooperation by virtue of network reciprocity. Naturally, this argument fails if leaders are made to conform because they then become unable to capitalize on their central position within the network. In the latter case, the coordination becomes less efficient, because players having the capacity to search for a more successful strategy are unable to lead others, which ultimately hinders the successful evolution of cooperation.

\begin{figure*}
\centerline{\epsfig{file=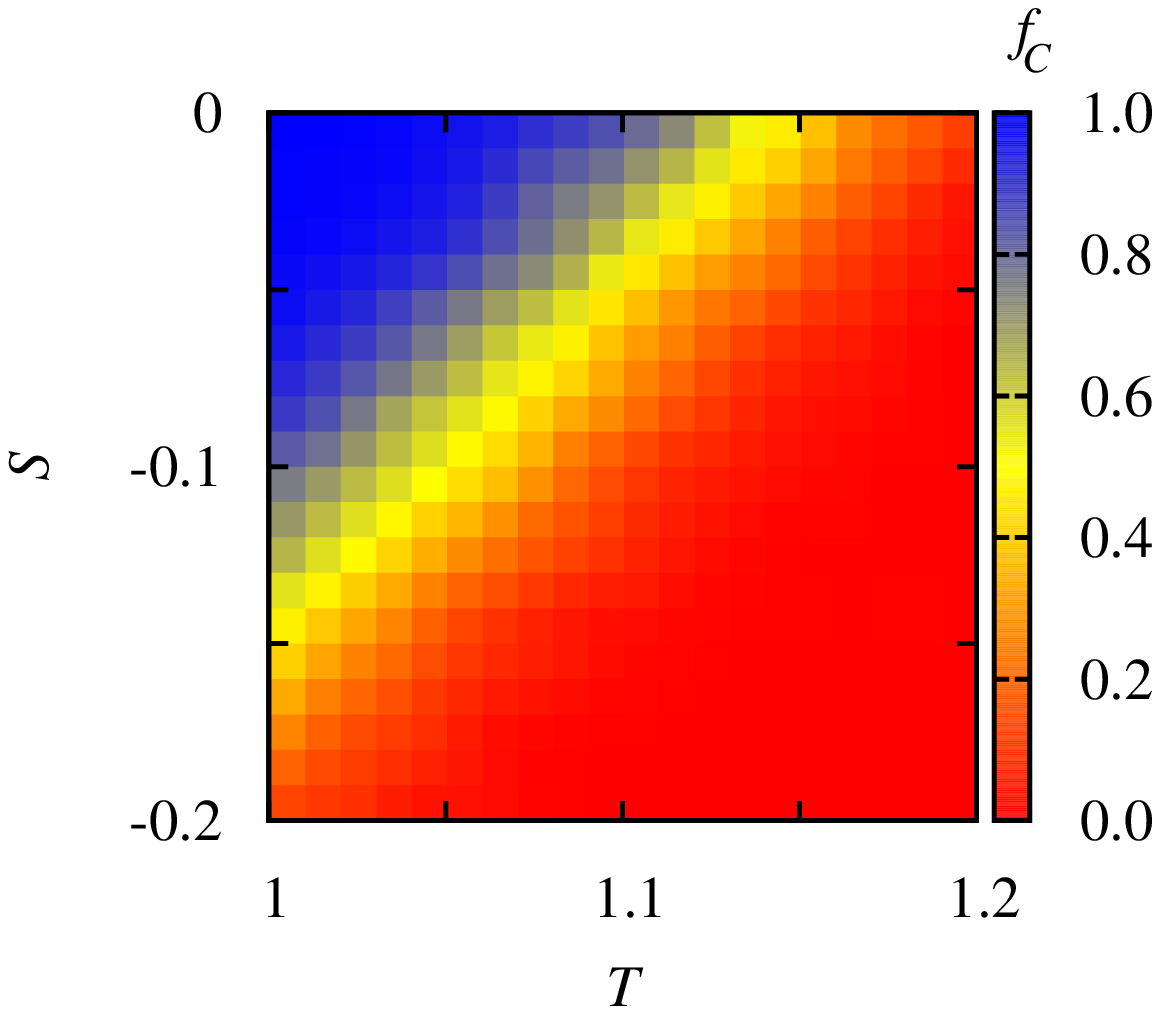,width=6.5cm}\hspace{1cm}\epsfig{file=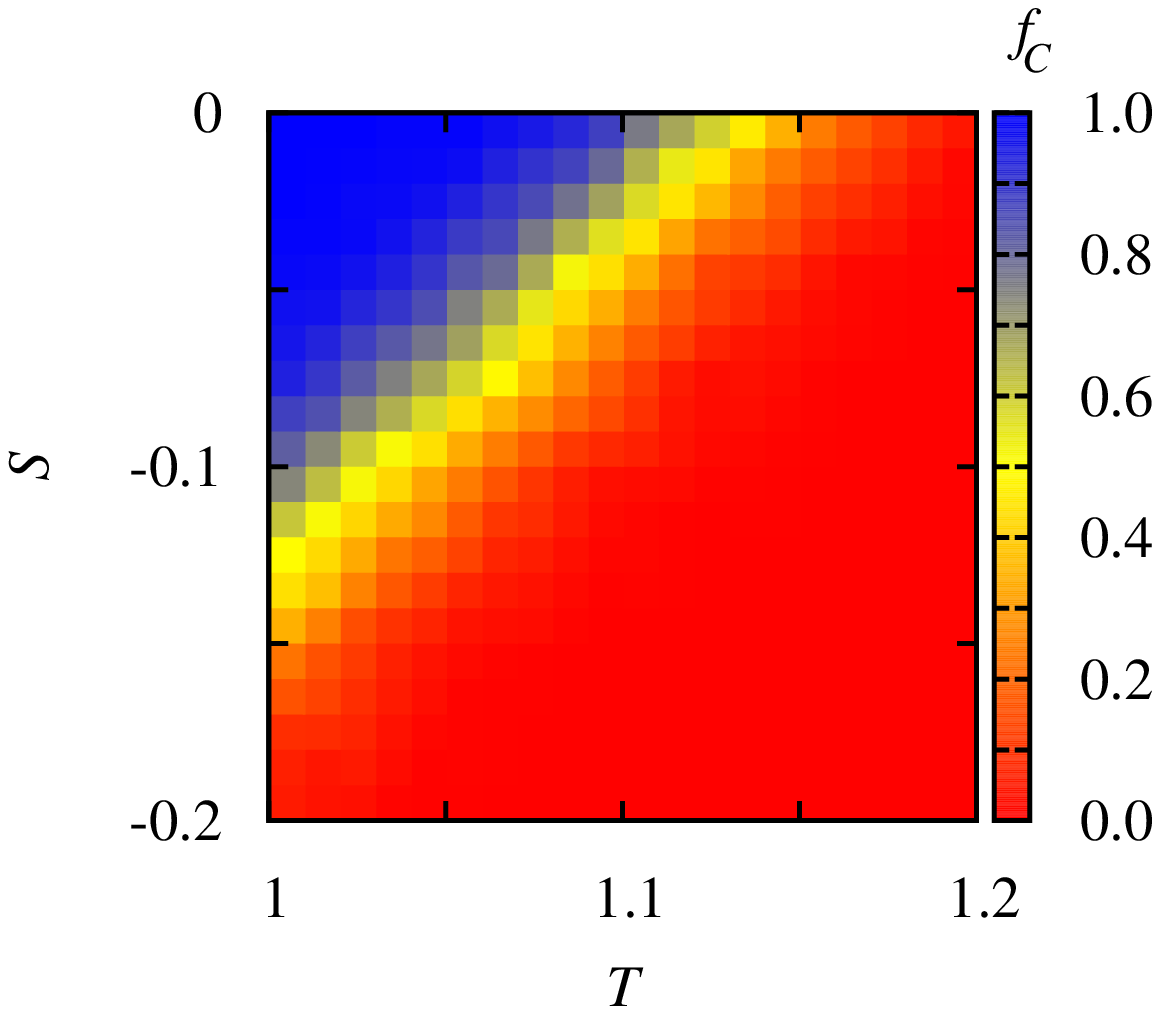,width=6.5cm}}
\caption{Color-coded fraction of cooperators $f_C$ on the $T-S$ parameter plane, as obtained when conformists are selected from low-degree players. Left panel shows the results obtained on scale-free networks, while the right panel shows the results obtained on Erd{\H o}s-R{\'e}nyi random networks. In both cases the fraction of the population that act as conformists was $\rho=0.7$. We note that, in the absence of conformists, and since we use degree-normalized payoffs, cooperation is practically absent in the depicted $T-S$ range. Networks with $N=10^4$ nodes have been used, and the presented results are averages over $1000$ independent realizations.}
\label{TS}
\end{figure*}

These details of the described macroscopic dynamics are generally valid and apply regardless of the properties of the interaction network and game parametrization. Indeed, as indicated by the results presented in Fig.~\ref{ER}, which were obtained on Erd{\H o}s-R{\'e}nyi random networks, it is better if conformists are selected specifically from those players who have lower degree or collective influence. On the contrary, if highly influential players are targeted as conformists and made to follow the majority, then the level of cooperation in the stationary state drops significantly. Similarly, as evidenced by the results presented in Fig.~\ref{TS} for both the scale-free (left panel) and Erd{\H o}s-R{\'e}nyi random (right panel) networks, even if the weak prisoner's dilemma game parametrization is replaced with the more stringent variant of the same social dilemma, the targeted selection of conformists amongst the masses still ensures high levels of cooperation in the stationary state. We note that, although the level of cooperation might appear to be modest at first, due to the application of degree-normalized payoffs, there would actually be almost no cooperators able to survive without the help of conformity in the considered $T-S$ range.

Previous research has also shown that, when we normalize payoffs on heterogeneous interaction network, we destroy the heterogeneity among players, which drastically weakens the positive impact of enhanced network reciprocity \cite{masuda_prsb07, tomassini_ijmpc07, szolnoki_pa08}. While this argument is of course valid, it is nevertheless important to keep in mind that players are still different because of the remaining differences in degree, and the resulting diverse opportunities to spread their strategies. This difference can be revealed if we choose conformists selectively, by considering the individual differences of players. As we have shown, it is detrimental to force leaders into following their own neighborhood. We note that this observation remains valid even if we apply less realistic absolute payoff values to determine strategy imitation probabilities \cite{pena_pre09}. On the contrary, expecting the same conformist attitude from the masses could be very useful for the whole population. This argument can be supported indirectly by considering also collective influence, as recently proposed in \cite{morone_n15}, instead of the degree. As can be observed in both Figs.~\ref{SF} and \ref{ER} (see label collective), exploiting the concept of collective influence to target the most influential players as conformists results in the same fall of cooperation as we have observed when high-degree players were made to conform.

\begin{figure}
\centerline{\epsfig{file=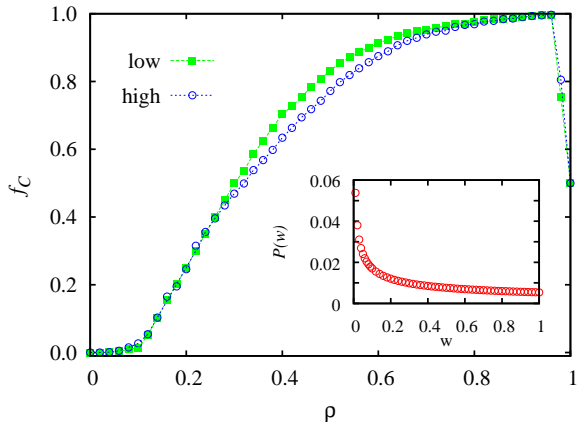,width=8.5cm}}
\caption{Fraction of cooperators $f_C$ in dependence on the fraction $\rho$ of the population that act as conformists, as obtained on the square lattice for two different selection rules indicated in the legend. As in Figs.~\ref{SF} and \ref{ER}, here assigning the conformist status to players with the highest strategy pass capacities (high) leads to lower levels of cooperation than are obtained if the same status is assigned to players with the lowest strategy pass capacities (low). Thus, even on lattices and regular networks (results for random regular graphs are practically identical), where the rank of a player is determined by a property not related to the topology of the network, it is clear that for socially optimal outcomes the conformists should be the masses, not the leaders. A square lattice with $N=800 \times 800$ nodes and the weak prisoner's dilemma $(T,S)=(1.1,0)$ parametrization have been used. Presented results are averages over $10$ independent realizations. The inset shows the applied distribution of strategy pass capacities in the population $P(w) \propto w^\frac{1}{2}$.}
\label{SQR}
\end{figure}

However, it is also possible to verify the above described conclusion directly if we distinguish players on regular networks or lattices, of course not by means of their degree or collective influence (because everybody has the same), but by means of their diverse capacities to pass strategy to the neighbors. In the latter case we apply a player specific prefactor $w$ to Eq.~\ref{fermi} which characterizes how successfully a player can deliver a strategy to the other player \cite{szolnoki_epl07}. To that effect, we assume that the strategy pass capacity distribution in the population is $P(w) \propto w^\frac{1}{2}$, where $w_x \in [0.01,1]$ interval, as show in the inset of Fig.~\ref{SQR}. This practically means that all players have a minimal chance to spread their strategy (hence we avoid frozen states), but only a small fraction of the population has $w_x \approx 1$. In this context the former players are the masses, while the later players are the more influential leaders. Evidently, we can select conformist preferably among the high-$w$ players (high in Fig.~\ref{SQR}) or among the low-$w$ players (low in Fig.~\ref{SQR}). As expected, for small and high $\rho$ values there is no relevant differences between the two selection rules. In the former case the impact of conformist players is negligible, while in the latter case it is too strong (a strategy neutral state is reached at $\rho=1$), either way precluding the observation of notable differences. In the intermediate region, however, we can see that higher cooperation levels can be obtained if conformity-driven players are chosen from low-$w$ group. In other words, it is better to allow leaders to search for higher individual payoffs, and in parallel, we should encourage the masses to conform. If we reverse these roles, leaders loose their ability to lead, which ultimately weakens network reciprocity.

\section*{Discussion}
We have studied the evolution of cooperation in social dilemmas where a fraction of the population aspires to conform rather than to maximize payoff. Thus, instead of choosing the strategy that is performing best in terms of payoff, conformists simply choose the strategy that is most prevalent in their interaction range. We have confirmed that, in general, conformity promotes cooperation by enhancing network reciprocity \cite{szolnoki_rsif15}. More precisely, because the dynamics between conformity-driven players is conceptually similar to the so-called majority-voter model \cite{oliveira_jsp92}, the interfaces between competing domains become smoother, which effectively minimizes invasion possibilities for the defectors.

What we have focused on specifically in the present paper is the fact that individuals within a population will typically be different in terms of the influence they are able to exert on others. The question then becomes who should conform if the goal is a socially desirable evolutionary outcome with ample cooperation. Should it be leaders, i.e., players with the highest degree, collective influence, or teaching activity, or should it be the masses, i.e., players with the lowest of these properties? Is it beneficial for the society if leaders tend to follow the masses? Using the simplest alternative, which is to assign conformists randomly regardless of their rank, as the benchmark, we have shown that leaders should not be conformists in evolutionary social dilemmas. Socially most favorable outcomes emerge only if the masses conform. This result is both intuitive as well as easily understandable in terms of the macroscopic dynamics that conditions it. In particular, we have argued that forcing the leaders to conform significantly hinders the constructive interplay between heterogeneity and coordination. Leaders must be able to create a following, which ultimately results in homogenous groups or clusters competing in the population. This competition is then swayed in favor of cooperators by the virtue of network reciprocity, which is further enhanced by smooth(er) interfaces that emerge due to the coordination of conformists. If, on the contrary, the leaders are made to follow, an unnatural coordination emerges -- the masses are expected to drive cooperation, but their limited interaction range and related weakness preclude this -- which ultimately hinders the successful evolution of cooperation.

Our results suggest that conformity might have had an evolutionary origin in that it promotes prosocial behavior. Although payoff maximization is surely an apt description of interactions among simpler forms of life, the preference to conform might also be a good heuristic in social decision making as it can help to avoid punishment or earn a reward \cite{rand_n12}, or it could also be the result of an effort to internalize the most common strategies among their neighbors. In the latter case, it could also be a source of an emerging social norm \cite{capraro_prsb15}. Regardless of what the goal of an interaction is, the present study is a simple example of ``multi-target'' evolutionary games, where the diversity of individual targets one may hope to achieve is mirrored in the array of strategies one can adopt for each specific purpose. This theoretical framework offers many possibilities to explore in the future, such as changing goals over time, considering not only expected payoffs but also the associated distribution of payoffs \cite{hagel_sr16}, or looking at the coevolution of behavior and cognition \cite{bear_pnas16}. We conclude with the hope that this paper will help motivate research along these lines in the near future. \\ \\

\section*{Methods}
As the interaction network, we use either homogeneous networks with a uniform degree distribution, such as the square lattice with periodic boundary conditions or random regular graphs with the same average degree $\langle k \rangle=4$, or heterogeneous graphs including
Erd{\H os}-R{\'e}nyi random networks with average degree $\langle k \rangle=6$, or Barab\'asi-Albert scale-free networks with an average degree $\langle k \rangle=4$. For the generation of the latter, we implement the standard growth and preferential attachment algorithm, as described in \cite{barabasi_s99}.

In terms of the simulation procedure, we note that each full Monte Carlo step (MCS) consists of $N$ elementary steps described above when introducing evolutionary games with conformists, which are repeated consecutively, thus giving a chance to every player to change its strategy once on average. All simulation results are obtained on networks typically comprising $N=10^4-10^5$ players. We determine the fraction of cooperators $f_C$ in the stationary state after a sufficiently long relaxation time lasting up to $10^5$ MCS. To further improve accuracy, the final results are averaged over up to $5000$ independent realizations, including the generation of scale-free and Erd{\H os}-R{\'e}nyi networks, as well as random initial strategy distributions, for each set of parameter values.

We also note that the results are independent of whether the assignment of conformity to the fraction $\rho$ of the population is done only once at the start of the simulation or if conformity is assigned to players at each instance of the game anew.

\begin{acknowledgments}
This research was supported by the Hungarian National Research Fund (Grant K-101490) and the Slovenian Research Agency (Grants J1-7009 and P5-0027).
\end{acknowledgments}

\end{document}